\documentclass[conference]{IEEEtran}
\usepackage{color,graphicx}
\usepackage{amsmath}
\usepackage{algorithmic}
\usepackage{algorithm}
\usepackage{epstopdf}
\usepackage{epsfig}
\usepackage{url}

\ifCLASSOPTIONcompsoc
  \usepackage[compress]{cite}
\else
  \usepackage{cite}
\fi

\ifCLASSINFOpdf

\else

\fi

\begin{document}
\title{Leveraging Decoupling in Enabling Energy Aware D2D Communications}

\author{\IEEEauthorblockN{Mukesh Kumar Giluka, M Sibgath Ali Khan, Vanlin Sathya and Antony Franklin A}
 \IEEEauthorblockA{Indian Institute of Technology, Hyderabad,\\
Email: [cs11p1002, ee13p0003, cs11p1003, antony.franklin]@iith.ac.in}
}

\maketitle

\begin{abstract}
Downlink/Uplink decoupling (DUDe) in LTE networks has caught the attention of researchers as it provides better uplink SINR and reduced power consumption per device due to decoupled connection of a device with the Macro (in downlink) and a small cell (in uplink). These characteristics of DUDe can be exploited to encourage more D2D communications in the network. This paper first proposes a model to estimate decoupling region within which a device is allowed to perform DUDe. Then, it formulates an equation to calculate the total power saved by devices due to decoupling. Finally, the extra area due to decoupling which can be used to enable D2D pairs is calculated. Simulation results are shown based on different simulation scenarios for different objectives for better understanding the idea proposed.

\end{abstract}

\IEEEpeerreviewmaketitle
\section{Introduction}
\label{intro}
The contemporary cellular traffic is diverse in nature. It is a combination of downlink traffic such as web browsing and file downloading, symmetric traffic such as social networking and gaming, and uplink traffic such as sensors and M2M traffic. The characteristics of devices generating such traffic is also different. In order to support these manifold traffic along with increasing spectral efficiency, cellular networks are turning to heterogeneous in nature. The heterogeneous cellular networks basically consist of different types of base stations viz. Macro, Pico, Micro and Femto. This infrastructure setup of cellular networks gives its users an option to connect to multiple base stations simultaneously. Association of a device with a base station mostly depends on load on the base station and received SINR (signal to noise and interference ratio) by a device with respect to the base station.

\begin{figure}[htb!]
 \epsfig{width=9cm,figure=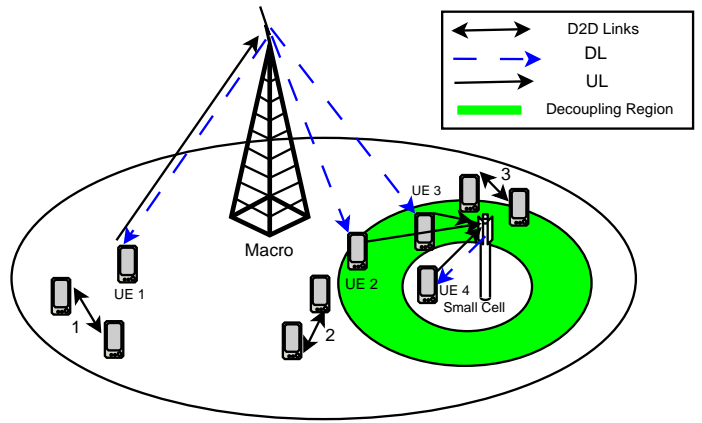}
 \caption{Typical LTE Network Scenario with D2D Pairs and Decoupling Devices.}
 \label{dude}
 \end{figure}

Apart from heterogeneity of cellular networks, different RAN improvement techniques are also being instrumental in enhancing the capacity of cellular networks. Two of these techniques are downlink/uplink decoupling (DUDe) and device-to-device communication (D2D). In DUDe~\cite{boccardi2015decouple}, a device is downlinkly connected to Macro and uplinkly connected to small cell. The decision of switching the base station by a device is taken based on uplink and downlink SINR of the device with respect to base stations. Figure~\ref{dude} shows a typical decoupling scenario where device UE1 and UE4 have coupled (both uplink and downlink connections with same base station) connection with Macro and small cell, respectively and UE2 and UE3 have decoupled connection having uplink with small cell and downlink with Macro. A device performs decoupling if it is placed in decoupling region. A decoupling region is the area within the coverage region of a Macro where a device receives better uplink SINR from small cell but better downlink SINR from Macro. In the figure, shaded region represents decoupling region. The main motivation to perform decoupling by a device is, increased uplink SINR and reduced power consumption. 

D2D~\cite{D2d} basically facilitates communication between devices in proximity. In D2D, control plane between two devices is established via eNodeB but having data communication between devices. Pair of devices having D2D connection is called as D2D pair. In Figure~\ref{dude}, pairs 1, 2, and 3 are D2D pairs. One of the important factor in enabling D2D between two devices is interference received by the receiver of the D2D pair. If a D2D pair and UE communicating via eNodeB, are assigned same resource blocks (RBs) then they will have to be sufficiently distanced from each other to avoid mutual interference during data transmission. 

In this paper, we have leveraged features of decoupling to enable more D2D pairs. Actually, when a device performs decoupling, it uses less transmit power to send its data in comparison to coupled connection. This reduction in power consumption can be used to enable more D2D pairs because interference caused by the decoupling device will be reduced if the D2D pair uses same RBs. Contributions of the paper are: (i) It first proposes a model to estimate the decoupling region. Decoupling region is basically the area with respect to a Macro and a small cell within which decoupling should be enabled. (ii) We have formulated an equation to calculate the total power saved by a mobile UE due to decoupling during its stay in the decoupling region and also, calculated the saved power by all static devices in the decoupling region by performing decoupling. And, (iii) We calculated the area within which more D2D pairs can be enabled if devices in the decoupling region follows decoupling. In best of our knowledge, this is the first work which utilizes features of decoupling to encourage D2D communication. Hereinafter, we have used DUDe and decoupling interchangeably for downlink/uplink decoupling. In this paper, we have used the words UE and device interchangeably.

Organization of the paper is as follows: Section~\ref{related_work} discusses works related to current state of the art in decoupling. Section~\ref{proposed} explains the proposed work. Section~\ref{simulation_results} presents simulation results and their analysis based on the idea proposed in previous section. Finally, Section~\ref{conclusion} concludes the paper. 
\section{Related Work}
\label{related_work}
In this section, we have presented the current state of the art in downlink/uplink decoupling. Paper~\cite{boccardi2015decouple} discusses  miscellaneous features of decoupling and shows the advantages of decoupling over coupled connection through simulations. In paper~\cite{elshaer2014downlink}, authors considered a static environment with single Macro cell and single small cell to compare the performance of DUDe with traditional coupled systems. Both non-interference and interference simulation environments are considered in the paper. In paper~\cite{smiljkovikj2015analysis}, authors analyse DL/UL association probabilities in the decoupling environment. They also studied the performance of decoupled access in a dense small cell deployment scenario while maximizing the average received power. They concluded that uplink throughput fairness will be there in case of decoupling. Paper~\cite{singh2014joint} presents an analytical model for a K-tier heterogeneous networks in which uplink SINR and rate coverage with load balancing considered. In paper~\cite{ants16}, authors presents a separate handover scheme for uplink and downlink in the DUDe environment. Apart from this, they analysed uplink SINR in a single Macro and multiple small cells interference scenarios for a mobile device. 

\section{Proposed Work}
\label{proposed}
This section is divided into three subsections: (i) Estimation of decoupling region, (ii) Analysis of power gain in decoupling scenario and (iii) Enabling D2D pairs through decoupling.
\subsection{Estimation of Decoupling Region}
As shown in Figure~\ref{dude}, green color shaded area shows the decoupling region. For each pair of adjacent base stations, there exists a decoupling region. 
In the decoupling region, following conditions must be satisfied:
\begin{itemize}
 \item Uplink SINR received by the small cell with respect to the device should be greater than that of Macro. In other word,
 \begin{equation}
 \label{uplinksinr}
 UplinkSINR_M < UplinkSINR_{S}
\end{equation}
\item Downlink SINR received by the device with respect to the Macro should be greater than that of small cell.
\begin{equation}
\label{downlinksinr}
 DownlinkSINR_M > DownlinkSINR_{S}
\end{equation}
 \end{itemize}
 
 If $P_{max}$ is the maximum power with which a device is allowed to transmit, $P_0$ is the target power which must be received by Macro or small cell, $K$ is the number of resource blocks assigned by the Macro or small cell to the device, $\alpha$ and $\beta$ are power control factors for Macro and small cell, $P_{T_M}$ and $P_{T_S}$ are transmit powers of the device with respect to Macro and small cell, $P_{L_M}$ and $P_{L_S}$ are path losses of the device with respect to Macro and small cell then following equations can be written:
\begin{equation*}
\label{power_equation11}
 P_{T_M}= \min(P_{max}, P_0log(K) + P_0 + \alpha P_{L_M})
\end{equation*}
\begin{equation*}
\label{power_equation12}
 P_{T_S}= \min(P_{max}, P_0log(K) + P_0 + \beta P_{L_S})
\end{equation*}
Since, $P_{max}$ is fixed for all devices and if we assume $K$ as one and $\alpha=\beta$ then the above equations can be written as follows:
\begin{equation*}
\label{power_equation21}
 P_{T_M}= P_0 + \alpha P_{L_M}
\end{equation*}
\begin{equation*}
\label{power_equation22}
 P_{T_S}= P_0 + \alpha P_{L_S}
\end{equation*}
If $P_{R_M}$ and $P_{R_S}$ are the received powers by Macro and small cell then
\begin{equation*}
\label{rpower_equation21}
 P_{R_M}= P_0 + (\alpha-1) P_{L_M}
\end{equation*}
\begin{equation*}
\label{rpower_equation22}
 P_{R_S}= P_0 + (\alpha-1) P_{L_S}
\end{equation*}
If we assume that there is one interferer present in the small cell region while the device is communicating with Macro then interference created by this interferer to Macro is given by:
\begin{equation*}
\label{interference_macro}
 I_M= P_0 + \alpha P_{L_S} - P_{L_{IM}}
\end{equation*}
where $P_{L_{IM}}$ is the path loss of the interferer with respect to Macro. Similarly, interference created by interferer, situated at Macro region, while device is communicating with small cell:
\begin{equation*}
\label{interference_smallcell}
 I_S= P_0 + \alpha P_{L_M} - P_{L_{IS}}
\end{equation*}
where $P_{L_{IS}}$ is the path loss of the interferer with respect to small cell. 
\begin{equation*}
\label{uplinkSINRM}
 UplinkSINR_M= \frac{P_{R_M}}{I_M+N_0}
\end{equation*}
Similarly,
\begin{equation*}
\label{uplinkSINRS}
 UplinkSINR_S= \frac{P_{R_S}}{I_S+N_0}
\end{equation*}
Since $I_M$ and $I_S$ are comparable and very much lesser than $P_{R_M}$ and $P_{R_S}$, values of $UplinkSINR_M$ and $UplinkSINR_S$ will mainly depend on $P_{R_M}$ and $P_{R_S}$. 

If $\rho_{G,M}$ and $\rho_{G,S}$ are the path gains of the device with respect to Macro and small cell, respectively, then $P_{R_M}$ and $P_{R_S}$ can be written as follows:
\begin{equation*}
\label{rpowergainmacro}
 P_{R_M}= \rho_0 \rho_{G,M}^{1-\alpha}
\end{equation*}
\begin{equation*}
\label{rpowergainsmallcell}
 P_{R_S}= \rho_0 \rho_{G,S}^{1-\alpha}
\end{equation*}
where, $\rho_0=10^{ \frac{P_0}{10}}$
Since the path gain depends on distance of the device from base station and $d_M$ and $d_S$ are distances of the device from Macro and small cell then the above equations can be written as follows:
\begin{equation}
\label{rpowergainmacrodm}
 P_{R_M}= 10^{\frac{P_0}{10}(\alpha-1)(35+30*\log \left(d_M\right))}
 \end{equation}
 \begin{equation}
 \label{rpowergainsmallcellds}
 P_{R_S}= 10^{\frac{P_0}{10}(\alpha-1)(35+30*\log \left(d_S\right))}
 \end{equation}
For Equation~\ref{uplinksinr} to be satisfied, value of $P_{R_M}$ must be lesser than $P_{R_S}$. This is possible if following condition is satisfied:
 \begin{equation}
  \label{distance_compare}
  d_M>d_S
 \end{equation}

If the co-ordinates of the Macro and the small cell are ($X_M$,$Y_M$) and ($X_S$,$Y_S$), respectively then any point ($x$,$y$) which satisfies above equation can be written as follows:
\begin{equation*}
\label{distance_square}
 (x-X_M)^2 + (y-Y_M)^2 > (x-X_S)^2 + (y-Y_S)^2
\end{equation*}
After simplifying, we can write
\begin{equation}
\scriptsize
\label{UL_decoupling_curve}
 x(X_M-X_S) + y(Y_M-Y_S)-\frac{(X_M)^2 - (X_S)^2+ (Y_M)^2-(Y_S)^2}{2} > 0
\end{equation}
Area under the curve denoted by the above equation is the region where Equation~\ref{uplinksinr} will be satisfied.
In case of Equation~\ref{downlinksinr}, in order to keep it simple, assume that only Macro and single small cell is there and the value of noise factor $N_O$ is zero then after putting the equations of downlink SINR of the device with respect to the Macro and the small cell and after simplification, it can be written that 
\begin{equation*}
\label{distance2}
 d_M > Kd_{S}
\end{equation*}
where K is a constant and its value is always greater than 1.
After simplifying the above equation in terms of ($x$,$y$), ($X_M$,$Y_M$) and ($X_S$,$Y_S$), it can be written as follow:
\begin{equation}
\scriptsize
\label{DL_decoupling_curve}
 x^2+y^2-\frac{2x(X_M-K^2X_S)}{1-K^2}-\frac{2y(Y_M-K^2Y_S)}{1-K^2}+\frac{Y_M^2-K^2Y_S^2}{1-K^2}>0
\end{equation}
Area under the curves denoted by Equations~\ref{UL_decoupling_curve} and~\ref{DL_decoupling_curve} is the decoupling area or region with respect to the Macro and the small cell.

\subsection{Analysis of Power Save in Decoupling Scenario}
As discussed in Section~\ref{intro}, in decoupling region, devices are associated with small cell for UL transmission and with Macro for DL transmission. 

From Equations~\ref{rpowergainmacrodm} and \ref{rpowergainsmallcellds}, we can write the following equations:
\begin{equation}
\label{power_equation41}
 P_{T_M}= 10^{\frac{P_0}{10}\alpha(35+30*\log \left(d_M\right))}
 \end{equation}
 \begin{equation}
 \label{power_equation42}
 P_{T_S}= 10^{\frac{P_0}{10}\alpha(35+30*\log \left(d_S\right))}
 \end{equation}
From Equations~\ref{power_equation41} and \ref{power_equation42}, it can be written that:
\begin{equation}
\label{powerfinal}
 P_{T_S} = (\frac{d_S}{d_{M}})^{3\alpha P_0}* P_{T_M}
\end{equation}
Since, in decoupling region $d_S<d_M$
\begin{equation*}
\label{power}
 P_{T_S} < P_{T_M}
\end{equation*}
From Equations~\ref{power_equation41}, \ref{power_equation42} and \ref{powerfinal}, we can say that power saved by an UE due to decoupling if UL transmission takes place in decoupling region will be $\Delta P_S$ where
\begin{equation}
\label{powersave}
 \Delta P_S = P_{T_M}*({1-(\frac{d_S}{d_{M}})^{3\alpha P_0}})
\end{equation}
Let, Macro and small cell are located on a straight line and a device is moving with constant velocity $v$ from Macro region to small cell region via decoupling region following the same straight line. Distances of the device from the Macro and small cell are $d_M$ and $d_S$, respectively at the decoupling point (point located at the boundary of the decoupling region) ($d_M$,$d_S$) lying on the same straight line. If the UE transmits in each $t$ seconds and there are $n$ number of transmissions performed by the UE in its stay in the decoupling region and if its first transmission occurs at the decoupling point ($d_M$,$d_S$) then total power saved by the device because of having a decoupled connection in the decoupling region will be:
\begin{equation}
\label{totalpowersave}
 \sum_{i=1}^n P_{T_{M,i}}*({1-(\frac{d_S-(i-1)vt}{d_{M}+(i-1)vt})^{3\alpha P_0}})
\end{equation}
where, $P_{T_{M,i}}$ is the transmit power of the device on its $i^{th}$ transmission to the Macro. $P_{T_{M,i}}$ can be written as follows:
\begin{equation}
\label{power_equationith}
 P_{T_{M,i}}= 10^{\frac{P_0}{10}\alpha(35+30*\log \left(d_M+(i-1)vt\right))}
 \end{equation}
If there are $m$ number of static devices located in the decoupling region having one time UL transmission and location of $j^{th}$ device with respect to Macro and small cell is ($d_{M_j}$,$d_{S_j}$) then total power saved by the total number of devices can be written as follows:
\begin{equation}
\label{totalpowersavedevices}
 \sum_{j=1}^m P_{(T,M)_{j}}*({1-(\frac {d_{S_j}}{d_{M_j}})^{3\alpha P_0}})
\end{equation}
where $P_{(T,M)_{j}}$ is the transmit power of the $j^{th}$ device with respect to Macro.
\begin{equation}
\label{power_equationithdevice}
 P_{(T,M)_{j}}= 10^{\frac{P_0}{10}\alpha(35+30*\log \left(d_{M_j}\right))}
 \end{equation}

\subsection{Enabling D2D Through Decoupling}\label{d2d_section}
While enabling D2D pairs, one of the important factor to be considered is interference from nearby devices (due to their transmit powers) using same resource blocks. If interference received by the receiver participating in D2D is below a threshold then D2D link can be enabled otherwise not. Decoupling can be helpful in enabling more D2D pairs in decoupling region. As clear from Equation~\ref{powersave}, the transmit power of a device can be reduced by following decoupling. This power reduction will cause reduction in interference received by devices planning for D2D and may enable them to perform D2D if interference goes below the threshold.
\begin{figure}[htb!]
 \epsfig{width=7cm, height=5cm, figure=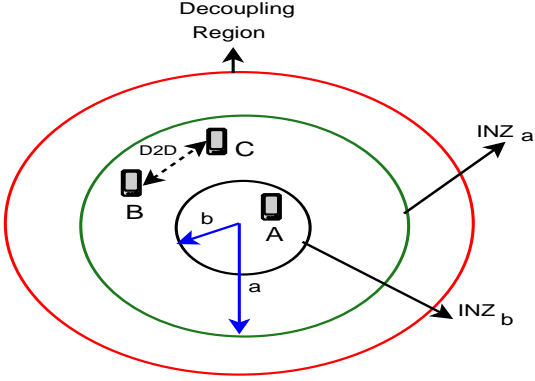}
 \caption{Interference Zones of D2D Pair with respect to a Decoupling Device.}
 \label{d2d_decoupling}
 \end{figure}

As shown in Figure~\ref{d2d_decoupling}, the outer boundary shows the decoupling region in which the device A is performing decoupling. The boundary with radius $a$ (denoted by $INZ_a$) and the boundary with radius $b$ (denoted by $INZ_b$) show the interference zones of device A when it was uplinkly attached with Macro and small cell, respectively. Here, the interference zone is meant by the area within which no D2D pair can be enabled due to interference caused by the power emitted by device A. Devices B and C want to become a D2D pair but due to lying in $INZ_a$, they are unable to form the D2D pair. But, due to decoupling, when device A connects with small cell in uplink, its interference zone changes to $INZ_b$ thereby devices B and C become eligible forming D2D pair. 

If $\lambda$ is the interference threshold (if desired D2D pairs receive interference more than this value then D2D can not be enabled), $P_{T_M}$ and $P_{T_S}$ are transmit powers of the device A with respect to Macro and small cell, then:
 \begin{equation}
 \label{d2dinza}
 \lambda= P_{T_M}+P_{L_a}
 \end{equation}
where $P_{L_a}$ is the path-loss of device A at distance $a$ when it is transmitting to Macro. Similarly,
\begin{equation}
 \label{d2dinzb}
 \lambda= P_{T_S}+P_{L_b}
 \end{equation}
where $P_{L_b}$ is the path-loss of device A at distance $b$ when it is transmitting to small cell after decoupling.\\
From Equations~\ref{power_equation41}, \ref{power_equation42}, \ref{d2dinza}, and \ref{d2dinzb}, we can conclude:
\begin{equation}
 \label{a_b}
 b=(a^{30}-10^{(35\alpha -1)}(d_M^{30\alpha} -d_S^{30\alpha}))^\frac{1}{30}
 \end{equation}

 
As clear from Equation~\ref{a_b}, the area of $INZ_b$ will be lesser than $INZ_a$. Hence, excess area ($\Delta A$) which can be used to enable more D2D pairs will be:
 \begin{equation}
 \label{excess_area}
 \Delta A= \pi(a^2-b^2)
 \end{equation}
 Let, there are $D$ number of devices communicating in the decoupling region. Let, for the $i^{th}$ device, interference zones with respect to Macro and small cell are $INZ_{a_i}$ and $INZ_{b_i}$, respectively where $a_i$ and $b_i$ are corresponding radii of interference zones. If we assume that interference zones of any two devices will never overlap and a D2D pair is under the interference of single decoupling device then total excess area ($\Delta A_T$) which can be used to enable more D2D pairs will be:
 \begin{equation}
\label{totalarea}
 \Delta A_T=\sum_{i=1}^D \pi({a_i}^2-{b_i}^2)
\end{equation}
where, 
 \begin{equation}
 \label{bivalue}
b_i=({a_i}^{30}-10^{(35\alpha -1)}(d_{M_i}^{30\alpha} -d_{S_i}^{30\alpha}))^\frac{1}{30}
 \end{equation}
 In the best case, if average number of D2D pairs per unit area can be enabled is $R$ then total number of extra D2D pairs can be enabled if devices perform decoupling in decoupling region is $R\Delta A_T$.

\section{Simulation Results and Analysis}
\label{simulation_results}
This section is divided into three subsections. Subsections show results related to issues discussed in Section~\ref{proposed} and follow the same order of presentation.
Results in all the subsections are taken in the scenario of single Macro and single small cell. We have used MATLAB for simulations.
\subsection{Time Spent in Decoupling Region}
\begin{figure*}[htb!]
\minipage{0.5\textwidth}
\epsfig{width=9cm,height=5cm,figure=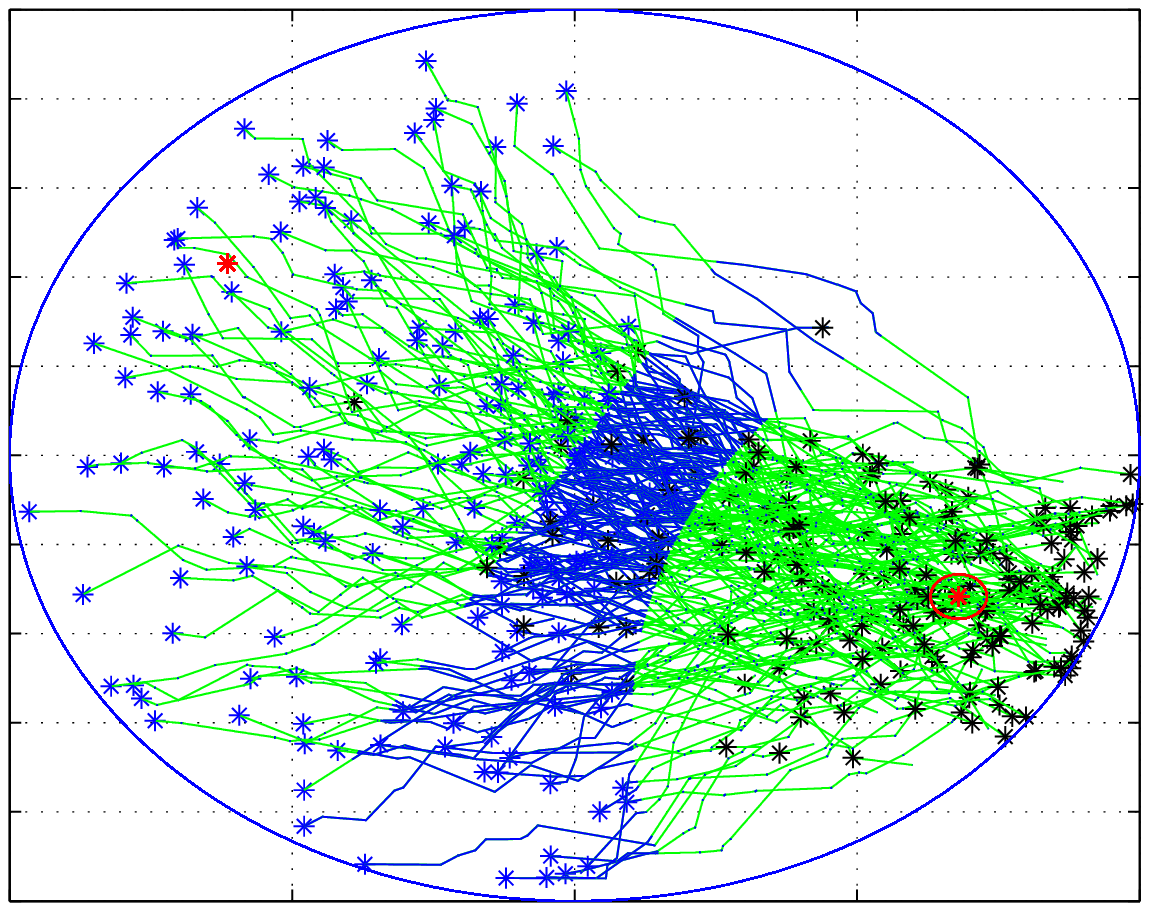}
 \caption{Decoupling Region.}
 \label{decoupling_region}
\endminipage\hfill
~
\minipage{0.5\textwidth}
 \epsfig{width=9cm,height=5cm,figure=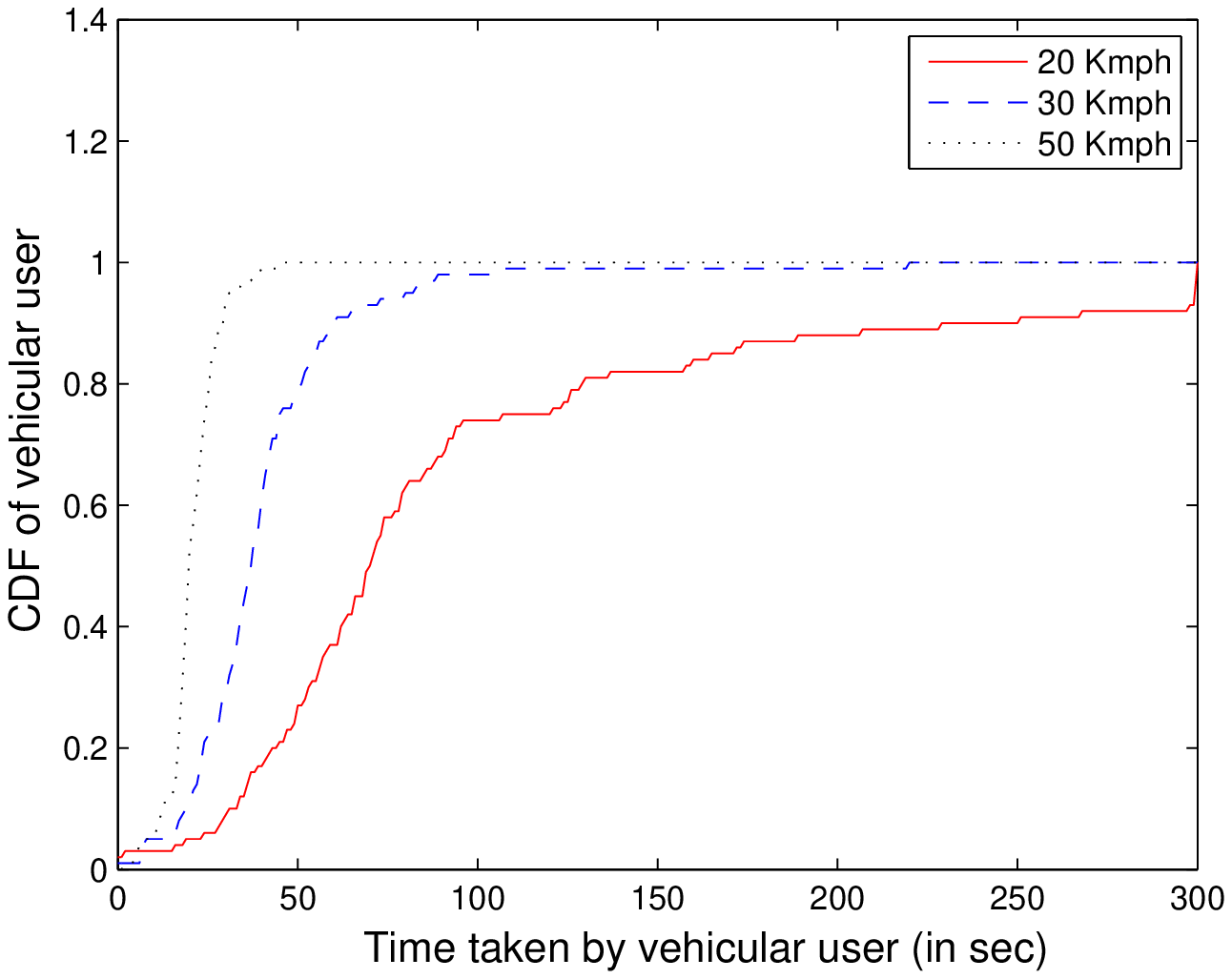}
 \caption{CDF of Decoupling Time of Devices Having Speeds 20, 30 and 50Kmph.}
 \label{veh1}
 \endminipage\hfill
 \end{figure*}

Here, we have shown and compared the cumulative distribution functions (CDFs) of decoupling time of devices in different mobility scenarios. Decoupling time is the time spent by a device in the decoupling region as calculated from Equations~\ref{UL_decoupling_curve} and~\ref{DL_decoupling_curve}. Figure~\ref{decoupling_region} shows a Macro and single small cell scenario where devices are moving from Macro region to small cell region. In the figure, the red color star in the left side denotes the position of the Macro and the red color star in the right side denotes position of the small cell. 
 There are 300 devices moving from the Macro region to the small cell region. The blue star shows their beginning point of the journey while black star shows the end point. Green lines are showing that devices are either in the Macro region or small cell region while blue lines are showing that devices are in decoupling region. The mobility model used for the simulation is Random Walk Mobility Model. Table~\ref{mobility_model} gives details of simulation parameters considered for the mobility model.

\begin{table}[htb!]
\caption{Simulation Parameters For Mobility Model}
\centering
\begin{tabular}{|p{1.7cm}| p{1.9cm}| p{3.6cm}|}
\hline\hline 
\bfseries{Parameter}&\bfseries{Model}&\bfseries{Model-Parameter}\\
\hline
Distance & Half Normal Distribution & Mean = 0.01 Km, Variance = 0.01 Km\\
\hline
Rotation Angle & Uniform Random & Range = $[\theta-\pi/4 , \theta+\pi/4]$ where, $\theta$ is angle between user and Femto cell with +ve x-axis.\\
\hline
\textbf{Velocity:} Vehicular & Half Normal Distribution & Mean = (20, 30, 50) Kmph, Variance = 10 Kmph\\
\hline 
\end{tabular} 
\label{mobility_model}
\end{table}

%
Figure~\ref{veh1} shows CDFs of time spent in the decoupling region by the devices. In Figure~\ref{veh1}, mean velocities are 20, 30, and 50 Kmph and 100 devices are associated with each velocity. Here, we can find that the time spent in the decoupling region is of the order of tens of second. In the figure, we can observe that as velocity of a device increases, its decoupling time decreases.

\begin{table}
\caption{Simulation Parameters}
\centering	
\begin{tabular}{|p{5.5cm}| p{2cm}|}
\hline\hline
\bfseries
\  Parameter&\bfseries Value \\ [0.2ex]
\hline	
Macro and small cell downlink transmit power & 40, 20 dBm \\
\hline
Maximum UE uplink transmit power  & 23 dBm \\
\hline
Number of RBs & 10 \\
\hline
Macro and small cell power control parameter ($\alpha$) and ($\beta$)  & 0.7, 0.7\\
\hline
Macro and small cell coverage radius & 1 Km, 0.035 Km \\
\hline
Scheduling algorithm & Round-Robin \\
\hline
\end{tabular}
\label{simulation}
\end{table}
\subsection{Comparison of SINR and Power Consumption in Coupled and Decoupled Scenarios}

\begin{figure}[htb!]
 \epsfig{width=8cm,figure=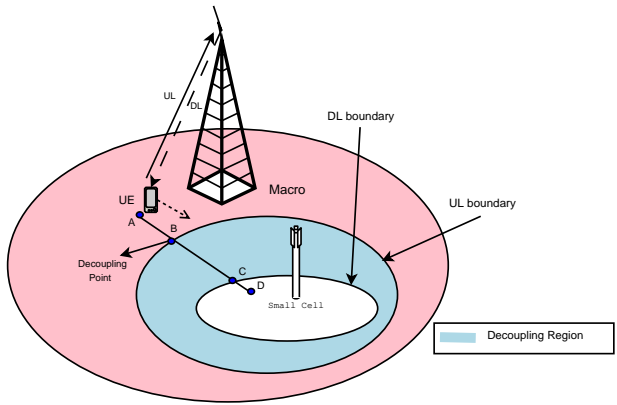}
 \caption{Decoupling Conditions due to Mobility of a Device.}
 \label{singlecell}
 \end{figure}

\begin{figure}[htb!]
\epsfig{width=9cm,figure=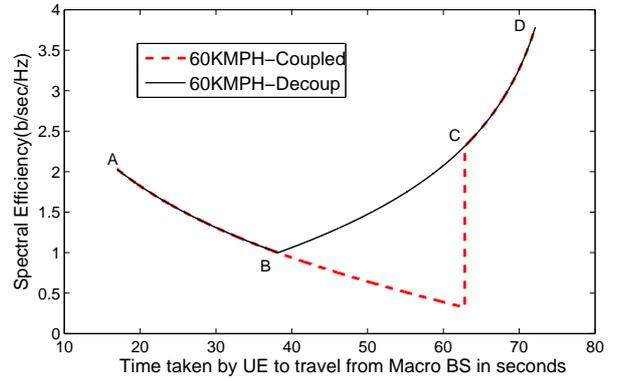}
 \caption{Spectral Efficiency Comparison for Coupling vs Decoupling in Mobility Scenario.}
 \label{single_speed}
\end{figure}

 Figure~\ref{singlecell} shows the scenario considered for simulation. The device moves from point 'A' to point 'D' and it has coupled connection with the Macro at point 'A', i.e., device is connected with Macro in both UL and DL. 'B' is the decoupling point for the device with respect to the small cell, i.e., device enters into the decoupling region at point 'B'. In the decoupling region, device will be connected to Macro in DL and with small cell in UL. At point 'C' device enters into the small cell region, hence, device will be connected to small cell in both DL and UL. As per the theoretical basis, in case of coupled connection, uplink SINR or spectral efficiency will keep decreasing from point 'B' to point 'C' and then increases from 'C' to 'D' while in case of decoupled connection uplink SINR will remain increasing from point 'B' to 'D'. 
 Table~\ref{simulation} shows simulation parameters considered for the results in this subsection.  Figure~\ref{single_speed} shows spectral efficiency of the device moving from point 'A' to 'D' with the speed of 60Kmph in both coupled and decoupled scenarios. Here, we can see that device's spectral efficiency starts increasing after point 'B' in decoupled scenario but for coupled scenario, the spectral efficiency decreases upto point 'C' and then increases.


\begin{figure}
\epsfig{width=9cm,figure=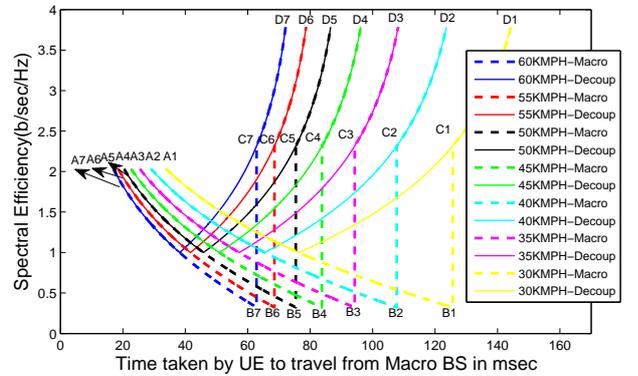}
\caption{Spectral Efficiency Comparison for Different Velocity.}
\label{multiple_speeds}
\end{figure}
 
\begin{figure*}[htb!]
\minipage{0.32\textwidth}
\epsfig{width=6cm,height=4cm,figure=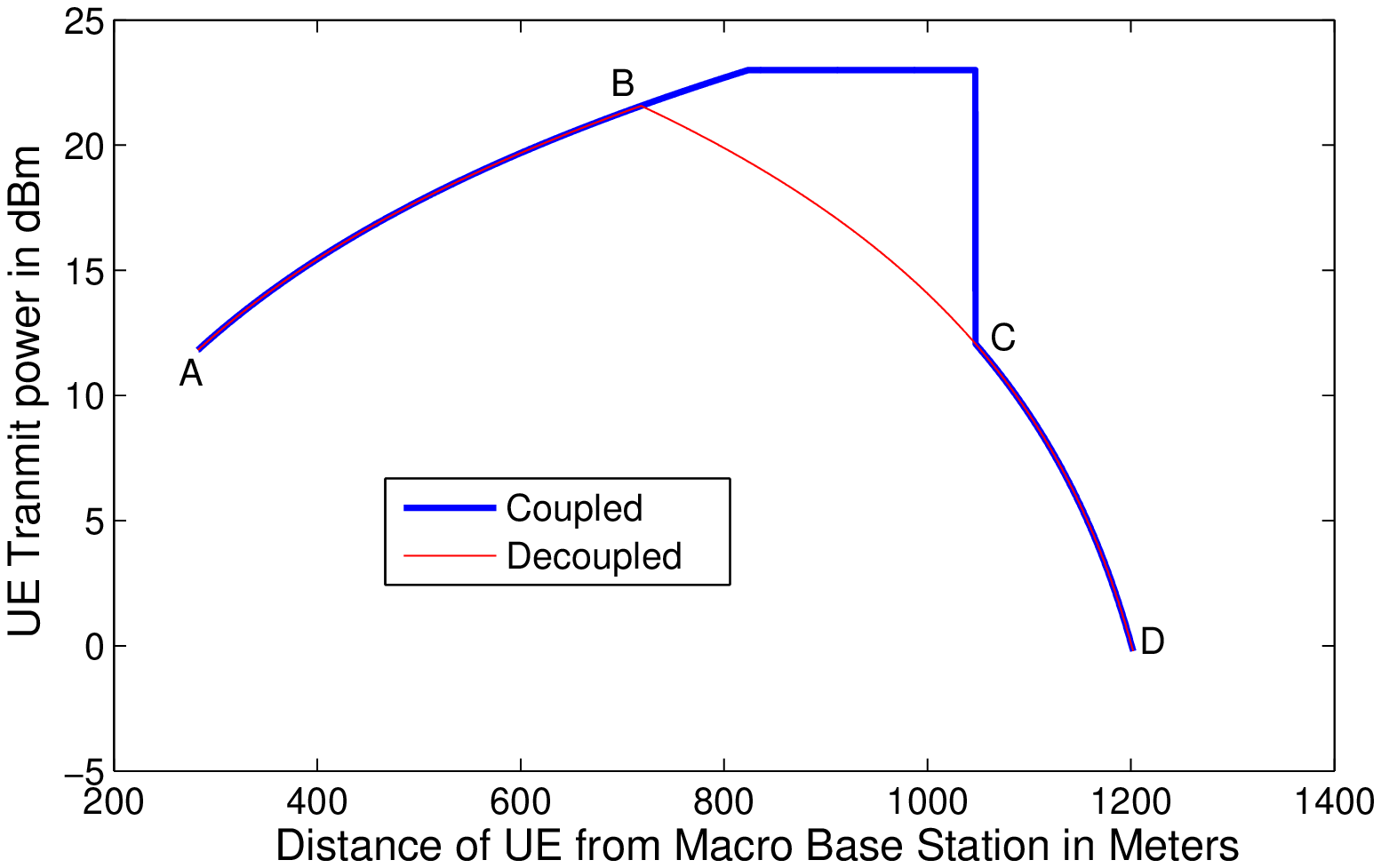}
 \caption{Comparison of Transmit Powers for Coupled and Decoupled Scenarios.}
 \label{transmit_power}
 \endminipage\hfill
 ~
\minipage{0.32\textwidth}  
\epsfig{width=6cm,figure=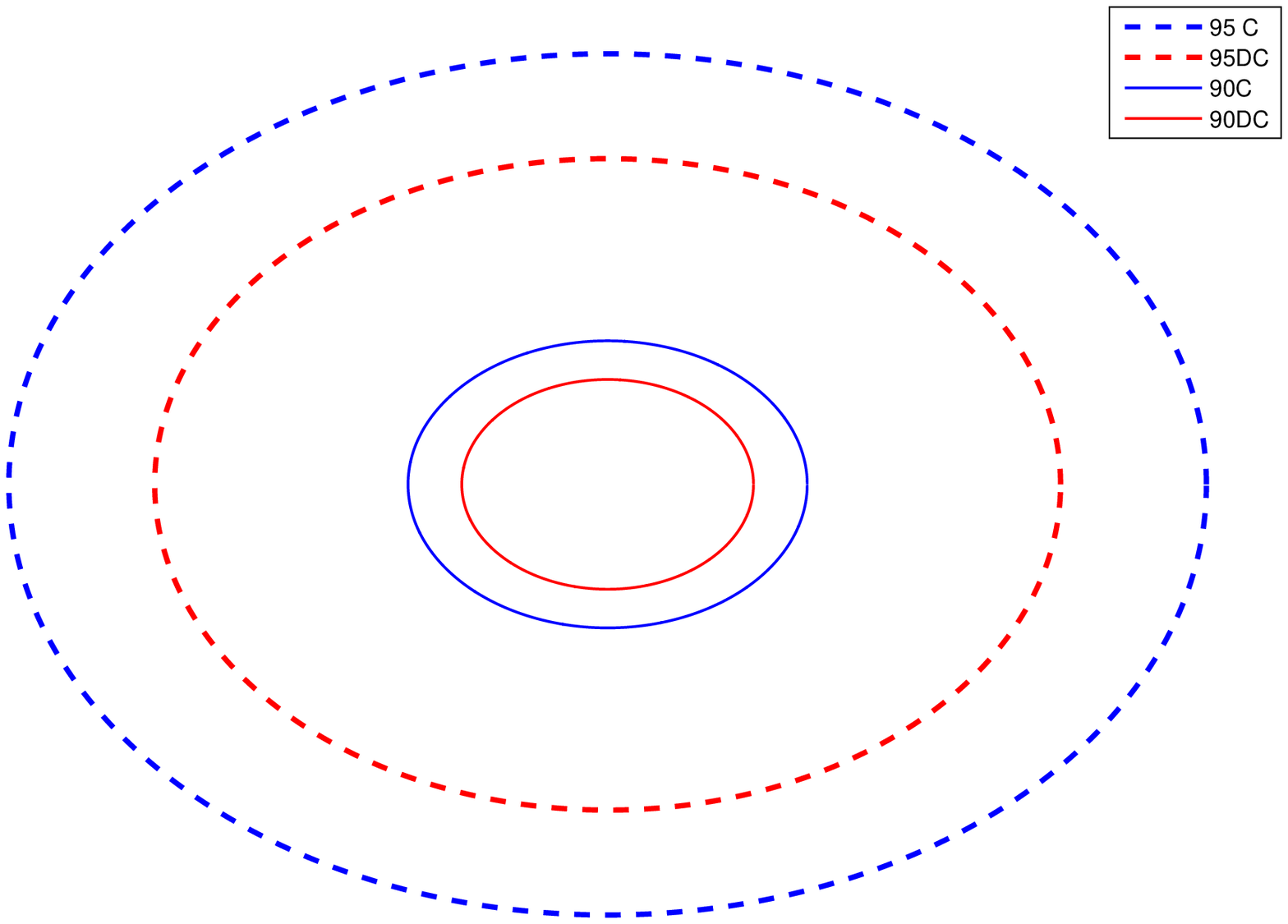}
 \caption{Interference Zone of a Decoupling Device Farther from Small Cell.}
 \label{Interf_90_95_60}
\endminipage\hfill
~
\minipage{0.32\textwidth}
\epsfig{width=6cm,figure=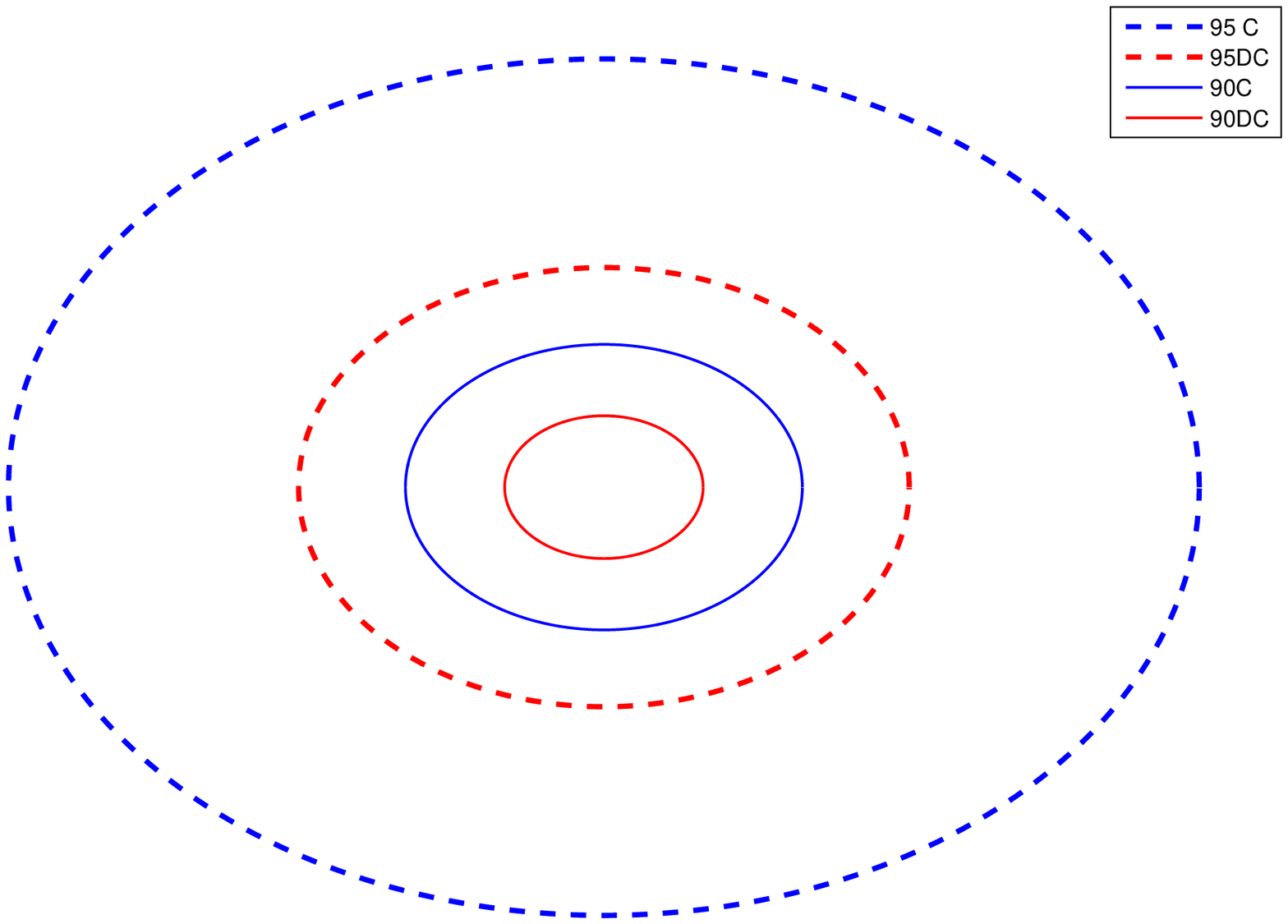}
 \caption{Interference Zone of a Decoupling Device Nearer to Small Cell.}
 \label{Interf_90_95_73}
 \endminipage\hfill
 \end{figure*}
 

 Figure~\ref{multiple_speeds} shows the spectral efficiency curves for different speeds of the device. $A_1, A_2, A_3, A_4, A_5, A_6$, and $A_7$ are starting points, $B_1, B_2, B_3, B_4, B_5, B_6$, and $B_7$, and $C_1, C_2, C_3, C_4, C_5, C_6$, and $C_7$ are decoupling points and $D_1, D_2, D_3, D_4, D_5, D_6$ and $D_7$ are end points of the movement of the device for speeds 30, 35, 40, 45, 50, 55 and 60, respectively. Geographically, $A_1$ to $A_7$ represents the same location which is point 'A' in the Figure~\ref{single_speed}. Similarly, $B_1$ to $B_7$, $C_1$ to $C_7$, and $D_1$ to $D_7$ represents points 'B', 'C', and 'D', respectively. In the figure, we can observe that as speed of the device increases, rate of decrement of spectral efficiency increases for coupled connection while rate of increment of spectral efficiency increases for decoupled connection. This happens because as the speed of the device increases, path loss in coupled connection (i.e., device and the Macro) increases and path loss in decoupled connection decreases with faster rate.
Figure~\ref{transmit_power} shows power consumed by a device to achieve uplink SINR of 0 dBm in both coupled and decoupled scenarios. Power consumption of the device increases as it moves away from the Macro but starts decreasing as it crosses the decoupling point 'B' in case of decoupling but continues to decrease in coupled scenario. After crossing point 'C', power consumption of the device starts decreasing in both scenarios.

\subsection{D2D and Decoupling}
In this subsection, we show through simulations that how decoupling can play role in enabling more D2D pairs. Table~\ref{table_d2d} presents the parameter values considered for simulations. 

We have compared the interference zones of a decoupling device for different values of interference thresholds of D2D pairs. As discussed in Section~\ref{d2d_section}, interference zone of a decoupling device is the area within which no D2D pair can be enabled due to interference caused by the decoupling device. Interference threshold of a D2D pair defines the interference tolerance capacity of the receiver of the D2D pair.

Figures~\ref{Interf_90_95_60} and \ref{Interf_90_95_73} show the interference zones of a decoupling device in case of both coupling and decoupling when the interference thresholds to enable a D2D pair are considered as 90 and 95 dBm, respectively. Legends 'C' and 'DC' are abbreviations of coupled and decoupled scenarios, respectively. These figures show that interference zones of a decoupling device is always smaller in case of decoupling, thereby, more D2D pairs can be enabled. In Figure~\ref{Interf_90_95_60}, the decoupling device is placed farther from the small cell while in Figure~\ref{Interf_90_95_73}, it is placed nearer to the small cell. In case of decoupling, distance of decoupling device from the small cell decides the area of corresponding interference zones. Nearer the decoupling device from the small cell, smaller is the interference zone. From Figures~\ref{Interf_90_95_60} and \ref{Interf_90_95_73}, we can conclude that with the help of decoupling, we can encourage enabling more D2D pairs. 

\begin{table}\caption{D2D Simulation Parameters}
\centering	
\begin{tabular}{|p{5.5cm}| p{2cm}|}
\hline\hline
\bfseries
\  Parameter&\bfseries Value \\ [0.2ex]
\hline	
Macro coverage radius & 800 Meters \\
\hline
Maximum transmit power of UE and D2D device  & 23 dBm \\
\hline
D2D interference thresholds & -90, -95 dBm \\
\hline
Distance of decoupling device from Macro  & 0.6, 0.73 Km\\
\hline
\end{tabular}
\label{table_d2d}
\end{table}
 \section{Conclusions}
 \label{conclusion}
 The key features of downlink/uplink decoupling are improved uplink SINR and reduced uplink power. In this paper, these features are theoretically modeled and further analysed through simulations. Apart from this,  we have used the power saving attribute of decoupling in interference minimization thereby enabling more D2D pairs. Decoupling devices nearer
to small cell have smaller interference zone and hence, able
to activate more D2D pairs.

\section*{ACKNOWLEDGMENT}\label{p4}

This work was supported by the Deity, Govt of India under the project Cyber Physical System (Grant No. 13(6)/2010CC\&BT).


\end{document}